\documentclass[12pt]{article}
\usepackage[english]{babel}
\usepackage{graphicx,rotating}

\makeatletter

\def\paragraph{\@startsection{paragraph}{4}{\z@}{+2.00ex plus
 +1ex minus +.2ex}{1.5ex plus .2ex}{\it\normalsize}}

\def\section{\@startsection {section}{1}{\z@}{+3.0ex plus +1ex minus
  +.2ex}{2.3ex plus .2ex}{\normalsize\bf\boldmath}}
\def\subsection{\@startsection{subsection}{2}{\z@}{+2.5ex plus +1ex
minus +.2ex}{1.5ex plus .2ex}{\normalsize\bf\boldmath}}
\def\subsubsection{\@startsection{subsubsection}{3}{\z@}{+3.25ex plus
 +1ex minus +.2ex}{1.5ex plus .2ex}{\normalsize\bf\boldmath}}

\expandafter\ifx\csname mathrm\endcsname\relax\def\mathrm#1{{\rm #1}}\fi

\newcounter{saveeqn}

\@addtoreset{equation}{section}

\def\ifmath#1{\relax\ifmmode #1\else $#1$\fi}%
\def\TeV{\ifmmode {\,\mathrm{ Te\kern -0.1em V}}\else
                   \textrm{Te\kern -0.1em V}\fi}%
\def\GeV{\ifmmode {\,\mathrm{ Ge\kern -0.1em V}}\else
                   \textrm{Ge\kern -0.1em V}\fi}%
\def\MeV{\ifmmode {\,\mathrm{ Me\kern -0.1em V}}\else
                   \textrm{Me\kern -0.1em V}\fi}%
\def\keV{\ifmmode {\,\mathrm{ ke\kern -0.1em V}}\else
                   \textrm{ke\kern -0.1em V}\fi}%
\def\eV{\ifmmode  {\,\mathrm{ e\kern -0.1em V}}\else
                   \textrm{e\kern -0.1em V}\fi}%

\newcommand{\ee}    {\mathrm{e}^+\mathrm{e}^-}

\newcommand{\bb}    {{\mathrm b\bar{\mathrm b}}}

\newcommand{\pp}    {\mathrm{p}\bar{\mathrm{p}}}

\newcommand{\ALR}    {A_{\mathrm{LR}}}
\newcommand{\Afb}    {A_{\mathrm{FB}}}
\newcommand{\Afbb}    {A_{\mathrm{FB}}^\mathrm{b}}
\newcommand{\Afbc}    {A_{\mathrm{FB}}^\mathrm{c}}

\newcommand {\bcl}   {\rm{b \rightarrow c \rightarrow \ell}}

\newcommand {\bl}   {\rm b \rightarrow \ell }

\newcommand{\MH}      {m_{\mathrm{H}}}

\newcommand{\MZ}      {m_{\mathrm{Z}}}
\newcommand{\MW}      {m_{\mathrm{W}}}
\newcommand{\MT}      {m_{\mathrm{t}}}

\newcommand{\GF}{G_{\mathrm{F}}}
\newcommand {\so}   {\sigma_0^{\rm{had}}}
\newcommand {\stl}  {\sin^2 \theta_{e\!f\!f}^l}

\newcommand {\cAe} {\mbox{$\cal A_{\rm e}$}}

\newcommand {\ahad}{\Delta \alpha_{\rm had}^{(5)}}

\begin{document}

\begin{titlepage}
\vspace*{1cm}
\begin{center}

{\Large \bf  Electroweak Precision Data and the Higgs Mass \\
 Workshop Summary}

\vspace*{1cm}

{\sc K. M\"onig}

\vspace*{.5cm}

{\normalsize \it
DESY-Zeuthen \\
Platanenallee 6, D-15738 Zeuthen}
\par
\end{center}
\vskip 1cm
\begin{center}
\bf Abstract
\end{center} 
{%\it
The status of the electroweak precision data as of winter 2003 and the
corresponding theoretical calculations are presented. The possible
problems within the data and the calculations are discussed in a
critical way. If the NuTeV anomaly cannot be explained by unknown effects
like higher order parton distribution functions and if it is not just
a statistical fluctuation the Standard Model of electroweak
interactions is in deep problems. Otherwise the conclusions that the
Higgs Boson is relatively light seems quite robust within the Standard Model.
}
\vfill
\begin{center}
{\it Invited talk presented at the Mini-Workshop\\
``ELECTROWEAK PRECISION DATA AND THE HIGGS MASS''\\
DESY-Zeuthen, February 28 to March 1 2003}
\end{center}
\end{titlepage}
%par
%vskip 1cm

\section{Introduction}
For a long time the electroweak precision data have been consistent with the
Standard Model prediction for a relatively light Higgs with fit
probabilities of order 50\% \cite{oldlepew}. With the more precise
data from LEP, SLD and the TEVATRON and especially with the $\sin^2
\theta$ measurement from NuTeV in deep inelastic neutrino nucleon
scattering the fit probability decreased to 1.3\% \cite{lepew02}. 
This decrease triggered the valid question if the
Standard Model is still valid and if we can still believe that the
Higgs is light.
 
\begin{figure}[htbp]
  \begin{center}
    \includegraphics[width=0.4\linewidth,bb=8 80 556 839]{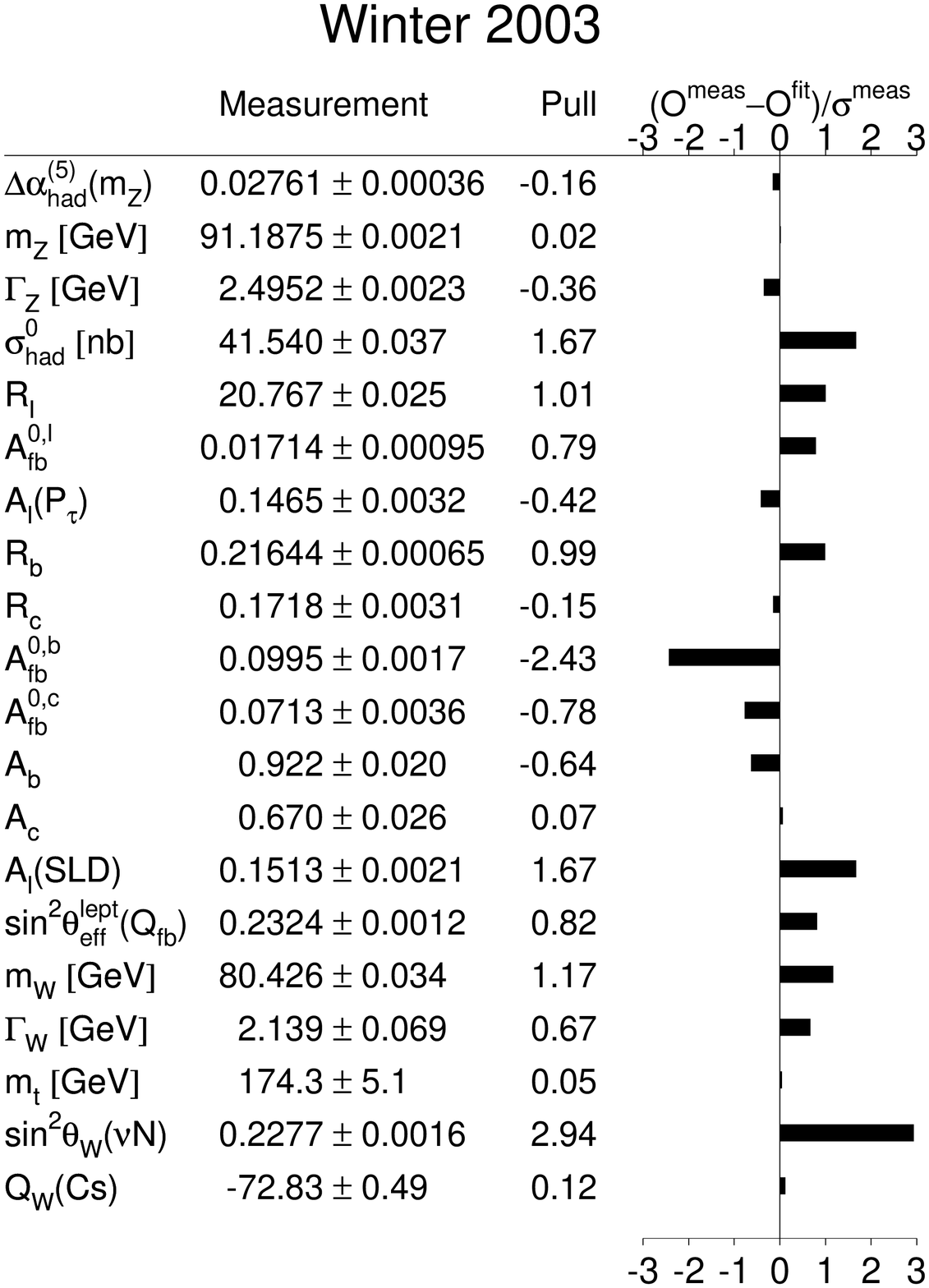}
    \includegraphics[width=0.57\linewidth,bb=18 38 561 565]{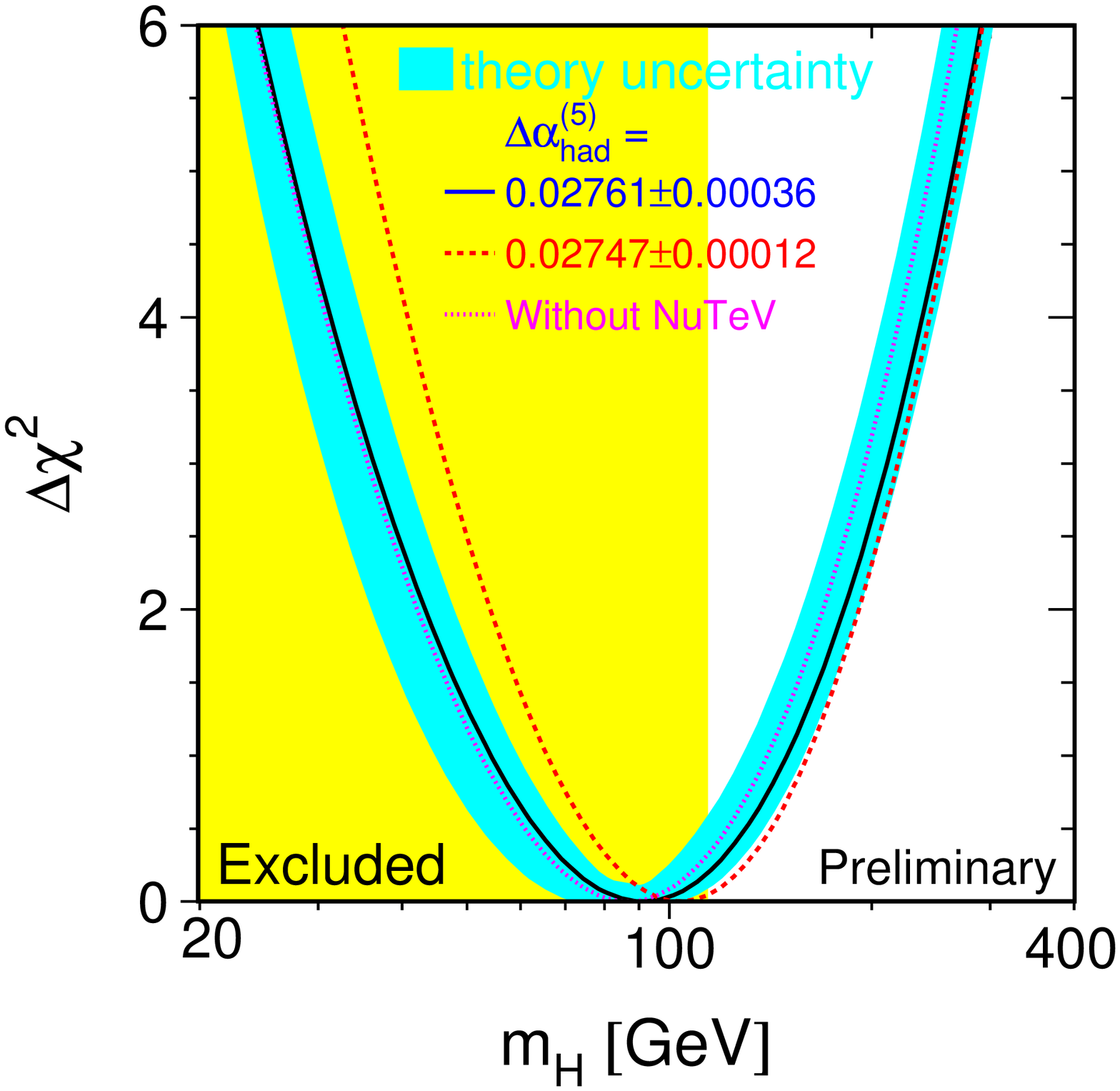}
    \caption{Left: Used data in the electroweak fit and their
    agreement with the Standard Model; right:
    $\Delta \chi^2$ as a function of the Higgs mass for the
      electroweak precision data.}
    \label{fig:blueband}
  \end{center}
\end{figure}

The workshop on ``Electroweak Precision Data and the Higgs Mass'' was
organised in order to answer the following questions:
\begin{itemize}
\item Can we believe the precision data?
\item Can we believe the theoretical calculations?
\item Are the data consistent within the model?
\end{itemize}
Within the workshop the data presented at the winter 2003 conferences
have been used \cite{martin}. As compared to summer 2002 the W-mass
has decreased by 0.7 standard deviations due to a correction of an
inadequacy in the ALEPH
calorimeter simulation. In addition new theoretical calculations
brought the atomic parity violation in Caesium from a $1.5 \sigma$
deviation exactly to the Standard Model prediction. These changes brought the
total $\chi^2$ of the fit to 25.5 for 15 degrees of freedom
corresponding to a probability of 4.4\%. 
The left plot in Figure \ref{fig:blueband} shows the data used in the
fit and their agreement with the fit prediction.
The largest deviations are
still the $\sin^2 \theta$ measurement from NuTeV with $2.9 \sigma$, the
b-quark forward-backward asymmetry from LEP, $\Afbb$, with $2.4 \sigma$ and 
the left-right asymmetry from SLD, $\ALR$, with $1.7 \sigma$. $\Afbb$
and $\ALR$ both measure the effective weak mixing angle $\stl$ and
deviate by roughly $3 \sigma$. How this deviation is distributed
between the pulls of the two measurements is determined by
the other Higgs mass dependent observables, mainly the W-mass, $\MW$.
The error breakdown for the worrying observables is shown in Table
\ref{tab:errbk}. $\MW$ has been included in this table because of its
correlation with the $\stl$ measurements. Its error breakdown is
approximate.
\begin{table}[htbp]
\begin{center}
\begin{tabular}[c]{|l|c|c|c|c|c|}
\hline
 & value & stat & exp. syst & theo. syst & pull\\
\hline
{$\sin^2 \theta (\nu N)$} 
& $0.2277$ & $0.0013$ & $0.0006$ & $0.0006$ & {$2.9$} \\
{$\Afbb$} & $0.0995$ & $0.0015$ & $0.0005$ & $0.0004$ & {$2.4$} \\
{$\ALR$} & $0.1513$ & $0.0018$ & $0.0010$ & $<0.0001\phantom{<}$ & {$1.7$} \\
{$\MW$} & $80.449$ & $0.024$ & $0.019$ & $0.017$ & {$1.2$}\\
\hline
\end{tabular}
\end{center}
\caption{Error breakdown of the worrying observables in the
 electroweak fit. The breakdown for $\MW$ is approximate and unofficial.}
\label{tab:errbk}
\end{table}

The global electroweak fit predicts the Higgs mass to be $\MH = 91
^{+58}_{-37} \GeV$. The $\Delta \chi^2$ as a function of the Higgs
mass is shown in the right plot of Figure \ref{fig:blueband}. 
Including theoretical
uncertainties, which are shown as the blue band, the 95\% c.l.~upper
limit on $\MH$ is $211 \GeV$.

\section{The Consistency of the Data with the direct Higgs Mass Limit}

The present situation has been outlined by Michael Chanowitz \cite{michael}. 
If all data are fitted the $\chi^2$ probability is only 1.9\%.%
\footnote{Details of the fit in \cite{michael} differ slightly from
  to the standard LEP Electroweak Working Group fit presented in
  \cite{martin}.} 
Even without the NuTeV result the probability
increases to only 17\%. In addition the central value of the Higgs
mass is somewhat lower than the LEP direct search limit decreasing the
combined probability, which is defined as the product of the $\chi^2$
probability and the probability that the Higgs is heavier than $114 \GeV$,
to 4.9\%. On the contrary, if one assumes
that the hadronic measurements of $\stl$ are not trustable, the fit
probability gets acceptable (71\%), however the combined probability
falls to about 3.5\% which is again worryingly low.

One should, however, notice, that the product of two flat probability
distributions is distributed like $- \ln (x)$ with a mean of 0.25, a
median of 0.18 and a most probable value of 0.
If instead a combined probability $P_c = P_1 P_2 (1 - \ln (P_1 P_2))$
is calculated, which is again flat between 0 and 1 if $P_1$ and $P_2$
are flat \cite{arb}, the combined  probability for the fit without
NuTeV is 20\%. For the fit without NuTeV and the hadronic $\stl$ \
measurements $P_c$ is 15\%.
Both probabilities don't appear worryingly low.
Using the procedure of the LEP Electroweak Working group, the
95\% c.l.~upper limit for the Higgs mass including theoretical
uncertainties in the last case becomes
$149\GeV$ sufficiently above the search limit.

\section{$\sin^2 \theta$ from NuTeV}

The $\sin^2 \theta$ measurement from NuTeV can be discussed
in isolation. The measurement contributes roughly nine units to the
$\chi^2$, but does not influence any of the fit results
significantly. In total there are three possibilities to interpret
this measurement. The measurement can be wrong or some theoretical
ingredient has been overlooked. In this case it is reasonable to
continue without it. The $3 \sigma$ can be just a statistical
fluctuation. In this case also all conclusions don't alter, only all
fit probabilities are worsened by this result. As a third possibility
the measurement is correct and the deviation from the prediction is
real. In this case the Standard Model breaks down in an unknown way
and the rest of this writeup becomes meaningless.

The progress in the NuTeV result comes from the separately available
neutrino and antineutrino beams \cite{kevin}. 
In this case $\sin^2 \theta$ can be
measured using the Paschos-Wolfenstein relation
\[
R^- = \frac{\sigma_{NC}^\nu - \sigma_{NC}^{\bar{\nu}}}
{\sigma_{CC}^\nu - \sigma_{CC}^{\bar{\nu}}} = 
\rho^2 \left( \frac{1}{2} - \sin^2 \theta \right)
\]
where $\sigma_{NC}$ ($\sigma_{CC}$) is the cross section for neutral
(charged) current interactions.
If the kinematic range of all four cross section measurements is the
same a lot of theoretical uncertainties cancel, especially the
dependence on the charm quark mass which was limiting this measurement
up to now.
In practice the charged to neutral current ratio is measured for
neutrino and antineutrino beams separately and $\sin^2 \theta$ is
fitted together with the charm quark mass using a Monte Carlo simulation.
Their final result, expressed as an on-shell mixing angle, is
\[
\sin^2 \theta (\nu N) = 1 - \frac{\MW^2}{\MZ^2} =  0.2277 
\pm 0.0013 ({\rm stat}) \pm 0.0009 ({\rm syst}).
\]
Experimentally the measurement looks solid and it will be assumed that
it is correct. Also most theoretical uncertainties have been checked,
however a few worries remain \cite{kevin}.

The analysis has been done with leading order parton distribution
functions (PDF). For the exact Paschos-Wolfenstein relation the dependence
on the parton distribution functions is almost negligible. However due
to the non equal kinematic range of the different measurements some
dependence might come in, so that it is worth to test the
Next-to-leading-order PDFs with errors. The
NuTeV result gets modified directly if the PDFs don't obey the assumed
symmetries, mainly the assumptions that the strange sea is symmetric
($s(x) = \bar{s}(x)$) and that there is isospin symmetry between the
proton and the neutron. A PDF analysis by Barone et al. \cite{barone}
suggests that a charge asymmetry in the strange sea exists with the
right size to explain the NuTeV deviation. This fit is however 
excluded now with the NuTeV data on charm production \cite{Goncharov:2001qe}.
The perturbative QCD corrections have been checked in NLO including
approximate experimental cuts and have been found to be rather small
\cite{kevin}. 

For isospin symmetry breaking the situation is not completely
clear. Typical models predict effects one order of magnitude smaller
than the experimental error, however the exact size of the violation
is hard to quantify.

As another possible explanation nuclear effects are discussed. The
anomaly could be explained if the nuclear effects are different in
charged and weak neutral currents. However the nuclear effects agree
well in charged and electromagnetic neutral currents, rendering this
possibility somewhat artificial.

In summary there is no hadronic effect that is a probable candidate to
explain the NuTeV anomaly. Nevertheless it is desirable that the data
are reanalysed with next to leading order PDFs 
using primarily NuTeV data.

Another possible worry are QED corrections. These corrections are
large and only one complete calculation exists \cite{bardin}.
In the  appropriate limits it agrees with calculations taking only
muon bremsstrahlung into account \cite{DeRujula:1979jj},
but nevertheless it would be good if the full 
calculation could be checked by another group.

\section{Consistency of the High Energy Data}

\begin{figure}[htb]
  \begin{center}
    \includegraphics[width=0.6\linewidth]{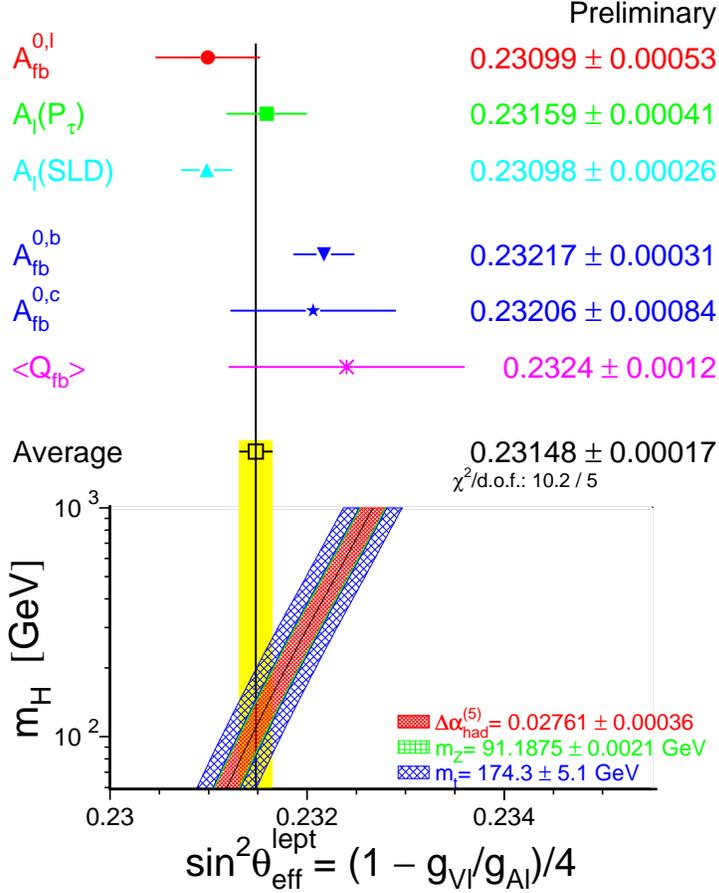}
    \caption{LEP end SLD measurements of the effective weak mixing angle.}
    \label{fig:sin2}
  \end{center}
\end{figure}

Since apart from the $\chi^2$ value the electroweak fit is not
affected by the NuTeV and the atomic parity violation data, in the
following only the high energy data will be discussed.
The most prominent inconsistency within the high energy data is the
$2.9 \sigma$ discrepancy between $\stl$ derived from $\ALR$ and
$\Afbb$. Figure \ref{fig:sin2} shows $\stl$ derived from the different
measurements at LEP and SLD. The average of all numbers yields
$\stl = 0.23206 \pm 0.00017$ with $\chi^2/ndf = 10.2/5$ corresponding
to 7\% probability. 

It is often argued that the hadronic measurements of $\stl$ cluster
around a high value while the leptonic ones are low. Table
\ref{tab:stl} shows the agreement of the other $\stl$ measurements
with $\ALR$ and $\Afbb$. If $\ALR$ is left out of the average the
$\chi^2$-probability is 36\%, if $\Afbb$ is left out, the probability
is 43\%. From this it can be concluded that apart from the
$\ALR-\Afbb$ discrepancy no further structure can be observed and that
the other measurements cannot decide which of the two might have a
problem.

\begin{table}[htbp]
\begin{center}
\begin{tabular}[c]{|l|c|c|}
\hline
observable & $\Delta \ALR [ \sigma ]$ & $\Delta \Afbb [ \sigma ]$\\
\hline
$\Afb^\ell$ & 0 & $-1.9$ \\
${\cal P}_\tau$ & $+1.3$ & $-1.1$ \\
$\Afbc$ & $+1.2$ & $-0.1$ \\
$<Q_{FB}>$ & $+1.2$ & $+0.2$ \\
\hline
\end{tabular}
\end{center}
\caption{Deviation of the other $\stl$ measurements from $\ALR$ and
  $\Afbb$ in units of standard deviations.}
\label{tab:stl}
\end{table}
The electroweak fit to the high energy data yields 
$\log(\MH) =  1.94 \pm 0.21$ with $\chi^2/{\rm ndf}  =  16.6/13$
corresponding to 22\% probability which is certainly not unacceptable.
The fit results, leaving out the worrying observables one by one are
shown in Table \ref{tab:fits}. Two conclusions can be drawn from these
fits. All fits have an acceptable $\chi^2$, however within the
Standard Model $\MW$ slightly prefers $\ALR$ compared to
$\Afbb$. But even if $\ALR$ and $\MW$ are excluded from the
electroweak fit the data prefer a relatively light Higgs.
As a further cross check the high energy data have been fitted
replacing all $\stl$ measurements by the LEP/SLD average 
$\stl = 0.23148 \pm 0.00012$. The fits give identical results to the
full one with $\chi^2/{\rm ndf}  =  6.5/8$.

\begin{table}[htbp]
\begin{center}
\begin{tabular}[c]{|l|c|c|c|}
\hline
left out & $\log(\MH)$ & $\chi^2/{\rm ndf}$ & Prob \\
\hline
{$\Afbb$} & $1.72$ & 9.2/12 & {69\%}\\
{$\ALR$} & $2.09$ & 13.1/12 & {36\%} \\
{$\MW$} & $2.00$ & 15.2/12 & {24\%} \\
{$\ALR$ \& $\MW$} & $2.24$ &  9.6/11 & {58\%} \\
\hline
\end{tabular}
\end{center}
\caption{Electroweak fits to the high energy data omitting some observables.
}
\label{tab:fits}
\end{table}

Theoretical uncertainties that effect derivations of specific
uncertainties, like the luminosity error or the error on QCD
corrections to the asymmetries are already included in the
uncertainties of the observables and thus in the $\chi^2$ definition.
On the contrary the uncertainties in the Standard Model predictions of
the pseudo observables are not accounted for in the $\chi^2$
definition and might therefore potentially affect the consistency of
the data. However, if as a representative test the recent corrections
to $\MW$ by Freitas et al \cite{freitas} that are implemented in
ZFITTER \cite{zfitter} are activated $\chi^2$ changes by only 0.3 and
if, as the authors of \cite{freitas} suggest, $\stl$ is increased
simultaneously by $8 \cdot 10^{-5}$, $\chi^2$ changes by
additional 0.5 so that the theoretical uncertainties on the prediction
of the pseudo-observables do not affect the consistency of the data.

\section{The Forward-Backward Asymmetry for b-Quarks}

To measure the forward-backward asymmetry for b-quarks at LEP ($\Afbb$)
\cite{pippa} three ingredients are needed. b-quark events need to be
identified, the quark direction needs to be measured and the quark
charge has to be tagged. For the quark direction the thrust axis is
always used. Thrust is infrared and collinear safe, so that QCD
corrections can be calculated and it is stable against hadronisation
effects. For the quark charge determination mainly two methods are in
use, leptons and jet-charge/vertex-charge techniques.
In the case of leptons one has to separate direct b-decays, $\bl$, from
cascade decays, $\bcl$, which lead to opposite sign leptons. The
separation is mainly done using the lepton momentum, $p$, and transverse
momentum with respect to the jet axis, $p_t$. These two variables can
also be used to identify $\bb$ events, so that in principle no other
flavour tagging algorithms are needed. However, some analyses use
lifetime tagging algorithms in addition to cleanup their samples.
For the jet-charge analyses the flavour tag always has to be done with
lifetime tagging. Because of the high efficiencies and purities of
these tags they can be calibrated mostly from data and only some small tagging
efficiencies for the background and hemisphere correlations have to be 
taken from the
simulation. The charge tag is a combination of jet-charge,
vertex-charge and possibly some additional information where either a
weighting method or a cut method is used. Details vary from experiment
to experiment but in all cases the charge tagging efficiencies are
calibrated from data comparing the charge assignment in the two
hemispheres.

Because of these self calibration procedures $\Afbb$ is largely
statistics dominated with a LEP-combined result of

\[
\Afbb = 0.0995  \pm   0.0015 {(\rm stat)} 
                \pm   0.0005 {(\rm exp \ syst)}
                \pm  0.0004  {(\rm cor \ syst)}.
\]

The largest correlated error in $\Afbb$ is due to QCD corrections
\cite{lephfqcd,neerven}. For full acceptance using the thrust axis as event 
direction the total correction is 
$\Afbb = \Afbb({\rm no \, QCD})  \cdot \left( 1- 0.0354 \pm 0.0063 \right)$.
This calculation contains all mass effects in first order and second
order for massless quarks. Two calculations exist using the quark
direction instead of the thrust direction \cite{afbneer,afbsey} which
agree numerically very well although some conceptual differences exist. 
One of them \cite{afbsey} exist also for the thrust axis and is thus used by
the LEP experiments. The error estimate contains uncertainties from
the knowledge of $\alpha_s$, higher order effects, quark mass effects 
and fragmentation and is considered to be conservative.

In the analyses b-quarks with high momentum are tagged
preferentially, so that the seen QCD corrections are typically a
factor two smaller. The bias factor is calculated with the simulation
and the applied correction and its error is scaled by this factor. The
uncertainty on the LEP-combined $\Afbb$ due to QCD corrections is 
$\Delta \Afbb ({\rm QCD}) = 0.00035$.

\section{$\stl$ from Polarised Asymmetries}

SLD measures $\stl$ mainly with the left-right cross section 
asymmetry \cite{morris}
\[
\ALR(\sqrt{s}) = \frac{1}{\cal P} 
\frac{\sigma_L - \sigma_R}{\sigma_L + \sigma_R}.
\]
Apart from Bhabha scattering, where also the t-channel contributes,
$\ALR$ is independent on the final state and, as long as the detector
acceptance is symmetric in the polar angle, independent from
experimental cuts. The cross section asymmetry at the SLC running
energy can thus be measured basically without systematic
uncertainties using a relatively tight hadronic event selection.
The largest challenge in the analyses is an accurate measurement of
the beam polarisation. With a Compton polarimeter behind the
interaction point and some other polarimeters for cross check, SLD was
able to measure the beam polarisation with a relative precision of
0.5\%.
The second largest systematic error source (0.4\%) is due to the
correction for $\gamma$Z-interference which depends strongly on the
beam energy. The error is mainly given from the statistics of a
miniscan to calibrate the energy spectrometer with the Z-mass.
The final result from the left-right asymmetry is
\[
\ALR^0  =  0.1514 \pm 0.0019 {\rm (stat)} \pm 0.0010 {\rm (syst)}
        =  0.1514 \pm 0.0022 
\]
which, after adding the polarised lepton asymmetries, results in
$\cAe = 0.1513 \pm 0.0021$. The measurement looks rather robust and
there are basically no theoretical uncertainties involved.

\section{The W-Boson Mass}

The W-boson mass, $\MW$, is measured at present with similar precision
in $\ee$ collisions at LEP and in $\pp$ at the TEVATRON. 

At LEP W-bosons are produced in pairs and practically all the
precision comes from reconstruction of the invariant mass of the
decaying Ws \cite{richard}. In roughly 45\% of the cases both Ws decay
hadronically and with the same probability one W decays hadronically
and one leptonically. The remaining events, where both Ws decay
leptonically, are not usable for the mass determination because of the
missing neutrinos. The resolution can be greatly improved by
constrained fits, forcing energy-momentum conservation. These fits
also reduce some systematic uncertainties, however they make the
W-mass dependent on the knowledge of the beam energy and initial state
radiation.
The experimental precision of the fully hadronic and the mixed decays is
roughly equal. However the fully hadronic events have a large
uncertainty from colour reconnection and Bose-Einstein correlations
between hadrons from the two Ws
which is hard to quantify. This uncertainty reduces the weight of the
fully hadronic events to about 10\% in the combination, so that it
plays no significant role in the final result. The LEP-combined W-mass
is
$\MW = 80.412 \pm 0.0029 ({\rm stat}) \pm 0.0031 ({\rm syst})
\GeV$. The largest systematic uncertainties are hadronisation ($18
\MeV$) and the knowledge of the beam energy ($17 \MeV$).

In $\pp$ collisions Ws are produced singly in quark-antiquark
annihilation \cite{baur}. Only leptonic W-decays can be used, since the
hadronic ones are completely hidden in the QCD background.
Traditionally the W-mass has been measured from the transverse
momentum spectrum of the decay lepton. This leads however to a large
uncertainty from the transverse momentum of the produced W. In the
current analyses the so called transverse mass 
$M_t = \sqrt{2 p_t^\ell p_t^\nu (1- \cos \varphi_{\ell \nu})}$
is used where the transverse momentum of the neutrino is reconstructed
from the lepton and the hadronic recoil. 
The main uncertainties are now the leptonic
energy scale and the hadronic recoil model. Both can be fixed with
leptonic Z decays assuming the Z-mass from LEP, so that the
corresponding uncertainties are mainly of statistical nature.

The largest theoretical uncertainties are from the parton distribution
functions and from QED effects. The PDF error ($15 \MeV$) is mainly
coming from events at the edge of the experimental acceptance. It
might increase slightly in future because the input errors in the PDF
fits could until recently not be properly propagated to the
results. On the other hand they should decrease in future because of
the better acceptance of the Run-II detectors and because they can be
constrained by TEVATRON data. The QED errors are $10-12 \MeV$ at
present, but improved calculations are under way. The combined W-mass
from the TEVATRON is $\MW = 80.454 \pm 0.059 \GeV$. The error is
systematics dominated, but the largest part of the systematic error is
of purely statistical nature.

\section{The Fine Structure Constant at the Z-scale}
\label{sec:alpha}

For the prediction of the precision observables the fine structure
constant at the Z-scale, $\alpha(\MZ^2)$, is needed. Its uncertainty is
mainly given by the contribution from the hadronic vacuum polarisation,
$\ahad$ \cite{fred}.

Several calculations of $\ahad$ exist that use in different ways the
cross section $\sigma( \ee \rightarrow {\rm hadrons})$
and perturbative QCD. Conservatively the
calculations which use data up to $\sqrt{s}=12\GeV$ are taken and the
value used by the LEP electroweak working group is the analysis from
Burkhardt and Pietrzyk \cite{pietrzyk}
($\ahad  =  0.02761 \pm 0.00036$) which leads to uncertainties of
$\Delta \stl  =  0.00013$ and $\Delta \MW  =  6.6 \MeV$ \cite{martin}.
This analysis uses the final results from BES in the $J/\Psi$ region
and preliminary data from CMD2 in the $\rho$ region. 
CMD2 corrected recently a bug in their normalisation \cite{cmderr}.
Using the final CMD2 data with the normalisation correction,
the Burkhardt and Pietrzyk analysis gives 
$\ahad  =  0.02768 \pm 0.00036$ \cite{pietrzyk2}.
An analysis from Jegerlehner, using the same data yields 
$\ahad  =  0.027773 \pm 0.000354$ in good agreement 
with Burkhardt and Pietrzyk \cite{fred}.
%It is however not completely clear if an analysis of Jegerlehner using
%the preliminary  CMD2 data agrees with Burkhardt and Pietrzyk, so that a
%possible bias in the analysis method should be investigated.

A problem at the moment is the disagreement between the
$\tau$ spectral functions and the CMD2 data which has been found in
the analysis of the hadronic contribution to $g-2$ \cite{davier}. 
The difference in
$\ahad$ using either the $\tau$ or the $\ee$ data without the normalisation 
correction corresponds to $0.8 \sigma$ of the used value. Adding the correction
it diminishes to about $0.6 \sigma$.
This discrepancy has to be resolved, especially for the understanding
of $g-2$. Some additional information might be obtained from radiative
return measurements at DA$\Phi$NE.
Adding the $\ee-\tau$ discrepancy as an additional error
to $\ahad$ increases the error on $\log \MH$ in the global fit
by less than 10\%.

The so called ``theory driven'' analyses, which use perturbative QCD
at lower energies decrease the error on $\ahad$ by up to a factor of
three. They all agree individually with the data driven ones, however
if they agree amongst each other is not completely clear. Because of
the smaller error it would, however, be desirable
that the differences could be understood.

\section{The Theoretical Error on the Luminosity}

The luminosity at $\ee$ machines is always measured using Bhabha
scattering at low angles and the precision critically depends on the
theoretical prediction of this process. Within the LEP/SLD 
pseudo-observables \cite{martin}
the luminosity error only affects the hadronic pole
cross section, $\so$. In the electroweak fit a change in $\so$ has
a modest effect on the strong coupling constant, $\alpha_s$. In
interpretations beyond the Standard Model it affects strongly the
number of light neutrino species, which is currently about $2\sigma$
below three.

Bhabha scattering at low angles has been modelled accurately and a
careful error estimate exists \cite{jadach}. The theoretical
uncertainty for the
analyses of the LEP experiments varies between 0.061\% and 0.052\%
roughly matching the experimental precision. The largest single
error source (0.04\%) is from hadronic vacuum polarisation. This
contribution can be reduced substantially with the new data from
CMD2, however, this suffers also from the $\ee - \tau$ discrepancy and
the error reductions requires that the central values of experimental
luminosities will change which makes a new combination of the results
necessary. 

\section{Pseudo Observables}

The experiments provide as results so called pseudo observables which 
are ``particle properties'' of the W and the Z, like their masses or
the ratio of the vector and axial vector coupling \cite{passarino}. 
To arrive at these
observables a three step procedure is needed. First the experiments
obtain their experimental signals which are per definition not
dependent on theoretical input. Out of these signals realistic
observables, i.e. cross sections and asymmetries within simple cuts
are constructed. This step introduces necessarily already some
dependence on QED and QCD. In the last step the pseudo observables are
obtained from these realistic observables which introduces further
dependence on QED and QCD but also requires corrections due to
$\gamma$-exchange, $\gamma$Z interference and imaginary parts of
couplings. 

It has been checked that this procedure is adequate at the present
level of accuracy. The theoretical uncertainties in the extraction of
the pseudo observables are typically less than one tenth of the
experimental errors. The experiments have performed fits of the Higgs
and top-mass and the strong coupling constant either directly to the
realistic observables or to the pseudo observables derived from
them. Both fits give identical results.

\section{Uncertainties in the Electroweak Predictions}

For the Higgs mass fit to the electroweak precision data the by far
most important observables are the effective weak mixing angle,
$\stl$, and the W-mass, $\MW$. The status of their predictions and
their theoretical uncertainties is discussed in detail in \cite{hollik}.
The derivation of $\MW$ from $\MZ$ and the Fermi constant $\GF$ if now
calculated completely in second order \cite{freitas,czakon} and the top quark
contributions in third order are known \cite{kuhn}. 
This leads to an uncertainty
in $\MW$ of about $3 \MeV$ if the input parameters are known. For
$\stl$ important two loop contributions are still missing. If one
takes the corresponding contributions to the W mass and varies the
on-shell mixing angle $\sin^2 \theta = 1- \frac{\MW^2}{\MZ^2}$ by this
amount one gets an error estimate of $\Delta \stl = 6 \cdot 10^{-5}$
which is almost half the experimental error. These uncertainties are included 
in the width of the blue band in Figure \ref{fig:blueband}.

\section{Fits within Supersymmetric Models}

Fits to the precision data have also been performed within
supersymmetric models \cite{weiglein}.
Since Supersymmetry with heavy superpartners looks identical to the
Standard Model with a light Higgs, Supersymmetry is clearly consistent
with the data. SUSY with light superpartners can fit the W-mass and
the anomalous magnetic moment of the muon, $g-2$, somewhat better than
the Standard Model, however the $g-2$ interpretation depends strongly on
the $\ee - \tau$ problem discussed in section \ref{sec:alpha}.
For $\sin^2\theta$ from NuTeV and the $\Afbb - \ALR$ discrepancy no
improvement can be achieved. In general the $\chi^2$ probabilities are
similar for the MSSM and the SM fits so that the electroweak precision
data cannot distinguish between the two models.

\section{Summary and Conclusions}

The electroweak precision data are consistent with the Standard Model
of electroweak interactions only on the 4\% level. The largest
deviation is the measurement of $\sin^2 \theta$ in neutrino nucleon
scattering with 2.9 standard deviations. This measurement is
relatively new and a large effort is still needed to understand it
theoretically in all details. However the deviation is still small
enough that it can well be a statistical fluctuation.

Apart from this measurement the agreement of the data with the
Standard Model fit is satisfactory and the prediction of a light Higgs
within this model seems rather robust. 
The second $3 \sigma$ effect, the disagreement between $\stl$ from
$\ALR$ and from $\Afbb$ does not spoil the fit quality and in the
probability for such a discrepancy to happen one has to take into
account that it is the largest difference selected out of many
possible combinations.
If any single measurement is excluded from the fit
the preferred Higgs mass stays rather low. As the first precision
observable the W-mass is
calculated fully to second order with the top mass dependent
corrections known in three loops leading to an error much smaller than the
experimental accuracy. For $\stl$ the theoretical uncertainty is about
half the experimental error making improvement in the prediction of
this quantity very desirable, but there seems no way how the $\stl$
prediction can
alter the Higgs mass conclusion. The same is true for $\alpha(\MZ^2)$
although also here it would be important to understand the problems
with this number.

In the near future two improvements on the experimental side can be
forseen. The error on the W-mass might shrink somewhat in the final LEP
analyses and will shrink substantially with the Run II data from the
Tevatron.
More importantly, also the error of the top-quark-mass from the
Tevatron will shrink by about a factor of two. At present the error on
$\stl$ and $\MW$ from the $\MT$ uncertainty is of the same 
size as the experimental errors and fully correlated between the two 
observables. This improvement also makes a better understanding of the
theoretical $\stl$ prediction and and $\alpha(\MZ^2)$ more important.

\end{document}